\documentclass[aps,prd,amssymb,showpacs,floatfix,preprint,nofootinbib]{revtex4-1}
\usepackage{amsmath,amsfonts,graphics,epsfig,amssymb,color}

\def\lsim{\raise0.3ex\hbox{$\;<$\kern-0.75em\raise-1.1ex\hbox{$\sim\;$}}}
\def\gsim{\raise0.3ex\hbox{$\;>$\kern-0.75em\raise-1.1ex\hbox{$\sim\;$}}}

\newcommand{\oleginline}[1]{{\color{black}#1}}

\newcommand\red{\color{red}}
\newcommand\black{\color{black}}
\renewcommand\red\black

\begin{document}
\title{Fluxes of diffuse gamma rays and neutrinos from cosmic-ray
interactions with circumgalactic gas}

\author{Oleg Kalashev}

\email{kalashev@ms2.inr.ac.ru}

\affiliation{Institute for Nuclear
Research of the Russian Academy of Sciences, 60th October Anniversary
Prospect 7a, Moscow 117312, Russia}

\author{Sergey Troitsky}

\email{st@ms2.inr.ac.ru}

\affiliation{Institute for Nuclear
Research of the Russian Academy of Sciences, 60th October Anniversary
Prospect 7a, Moscow 117312, Russia}



\begin{center}
\begin{abstract}
The Milky Way is surrounded by a gravitationally bound gas corona
extending up to the Galaxy's virial radius. Interactions of cosmic-ray
particles with this gas give rise to energetic secondary gamma rays and
neutrinos. We present a quantitative analysis of the neutrino and
gamma-ray fluxes from the corona of the Milky Way together with a combined
contribution of coronae of other galaxies. The high-energy neutrino flux
is insufficient to explain the IceCube results, while the contribution to
the FERMI-LAT diffuse gamma-ray flux is not negligible.
\end{abstract}
\end{center}
\maketitle

\section{Introduction}
\label{sec:intro}
The origin of recently discovered high-energy astrophysical neutrinos
\cite{IceCube1, IceCube2, HESE-3yr, muon-events}  is not obvious. The flux
of the neutrinos is quite large: they were observed already in two-year
IceCube data. The global distribution of arrival directions of the events
is consistent with isotropy, as expected for their extragalactic origin,
but the observed spectrum is a bit soft for powerful extragalactic
sources. Moreover, if the neutrinos are born in $\pi^{\pm}$-meson decays,
which is the most natural astrophysical scenario, the accompanying flux of
gamma rays from $\pi^{0}$ decays overshoots the observed isotropic diffuse
photon flux measured by FERMI LAT \cite{FERMIbckgr} below 820~GeV unless a
significant part of the neutrinos are of the Galactic origin. This is
because of electromagnetic cascades \cite{EMcascades, EMcascades1} which
transfer the energy of gamma rays from multi-TeV to sub-TeV bands by means
of efficient $e^{+}e^{-}$-pair production on the background radiation and
subsequent inverse Compton scattering. To overcome these troubles, models
with several components of the neutrino flux of completely different
origin, e.g.\ from cosmic-ray interactions with the interstellar matter
\textit{and} from distant blazars, are discussed. In these explanations,
two very different classes of sources give roughly equal contributions to
the observed neutrino flux by coincidence. Another option is to consider
optically thick sources which emit neutrinos but totally absorb photons.

An interesting proposal to explain the IceCube observation has been put
forward in Ref.~\cite{Ahar-gas}, where the interactions of cosmic-ray
protons escaping the Galaxy with the circumgalactic gas result in the
required diffuse neutrino flux. Note that a similar mechanism has been
proposed in Ref.~\cite{Gnedin} to contribute a significant amount to the
diffuse gamma-ray background at energies $\sim$GeV. The energy fluxes
carried by diffuse GeV photons and IceCube sub-PeV neutrinos are of the
same order, and one may hope that they might have a common origin.

In this
paper, we
calculate the corresponding
neutrino and gamma-ray fluxes, account for propagation of photons and
compare the resulting fluxes at the Earth with the observational data
obtained by IceCube and FERMI-LAT.
We demonstrate that the flux of secondary neutrinos from cosmic-ray
interactions with the circumgalactic gas is insufficient to explain
IceCube astrophysical neutrino events, if realistic density of
cosmic rays escaping the Galaxy,
normalized to the present-day cosmic-ray spectrum in the disk, is assumed.
At the same time, secondary photons from the same interactions contribute
substantially to the diffuse gamma-ray background at FERMI-LAT energies.

The rest of the paper is organized as follows.
In Sec.~\ref{sec:gas}, we briefly discuss observational data about the
Milky-Way circumgalactic gas and fix the gas density profile to be used in
our calculations. In Sec.~\ref{sec:CR}, we determine the density and the
spectrum of cosmic rays escaping from the Galaxy  and describe their
interactions with the gas. Section~\ref{sec:EG} addresses the contribution
of similar processes taking place in other galaxies throughout the
Universe. The results of our calculations are formulated and discussed in
Sec.~\ref{sec:concl}.

\section{The Milky Way's corona}
\label{sec:gas}
Recent observations suggest that
our Galaxy is surrounded by a huge (extending to $\sim
250$~kpc from the Galactic center) halo of gas which we call here the
Milky Way's corona
in order to clearly distinguish it from the halo
of stars whose size is approximately ten times smaller. Cosmic rays
escaping from the Galaxy interact with this gas; these interactions have
been considered as the source of (a part of) the cosmic diffuse gamma-ray
background \cite{Gnedin}. Order-of-magnitude
estimates in Ref.~\cite{Ahar-gas} suggested that the very same
interactions
may produce a significant part of the high-energy neutrino flux observed
by IceCube. The aim of the present paper is to calculate the corresponding
neutrino and gamma-ray fluxes in a more precise way.

There are two classes of observational results pointing to the existence
of the Milky-Way gaseous corona:
\begin{itemize}
\item
O{\small  \sc VII} and O{\small \sc VIII} absorption X-ray
lines observed at zero redshift in the spectra of extragalactic sources
and corresponding emission lines observed in the blank-sky spectrum, see
e.g.\ Refs.~\cite{gas1, MB_ApJ_770_118, 1412.3116};
\item
evidence for the ram-pressure stripping of MW satellite galaxies, see
e.g.\ Refs.~\cite{astro-ph/0001142, 0901.4975, 1305.4176, 1507.07935}.
\end{itemize}
X-ray spectroscopic observations are more precise statistically but have
large systematic uncertainties because the
oxygen is only a tracer of much more abundant hydrogen and other gases.
Derivation of the total gas density from these results depends crucially
on the assumptions about the metallicity $Z$ of the circumgalactic gas and
of the fraction $f$ of the particular observed oxygen ion. The most
precise studies~\cite{1412.3116} assume $f=0.5$ and $Z=0.3 Z_{\odot}$,
where $Z_{\odot}$ is the solar metallicity. These values are quite
arbitrary; moreover, they are assumed to be constant all the way up to the
virial radius, while qualitative arguments suggest that $Z$ should
decrease with the galactocentric distance $r$. In addition, recent
simulations \cite{1504.05594} and ultraviolet O{\small \sc VI}
observations \cite{1602.00689} suggest that the hot gas is accompanied by
a similar amount of cold and/or warm component, so that X-ray observations
may underestimate the total amount of matter by a factor of two.
Therefore, both the normalization and the radial dependence of the gas
density inferred from the X-ray spectroscopy remain largely uncertain.
Contrary, estimates of the gas density from ram-pressure stripping of MW
satellites are considerably less precise, but at the same time less
model-dependent.

\begin{figure}
\centerline{\includegraphics[width=0.95\columnwidth]{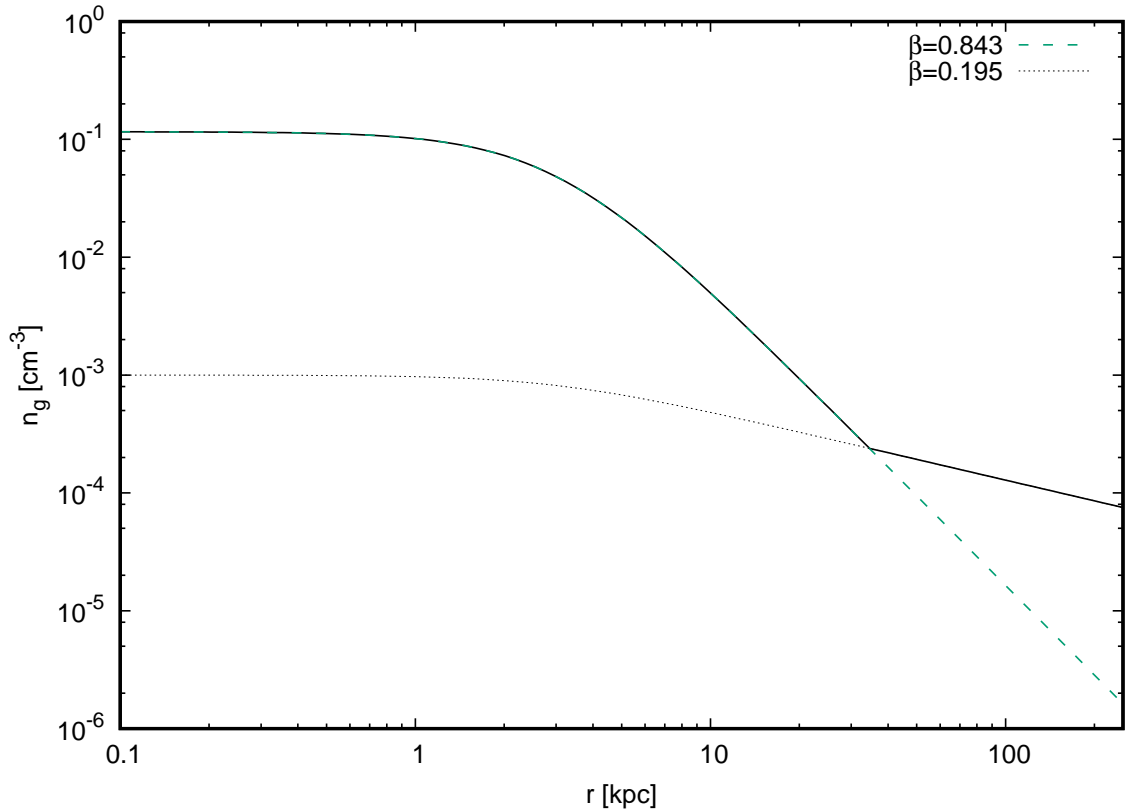}}
\caption{
\label{fig:ngas}
The circumgalactic gas density profile. Dashed: approximation of the
simulated profile of Ref.~\cite{Gnedin} valid at $r \lesssim 45$~kpc.
Dotted: the best-fit profile of Ref.~\cite{ST} valid at $r \gtrsim
45$~kpc. Full line: the profile used in this work.}
\end{figure}

In Ref.~\cite{ST}, the two classes of observations are combined to
constrain the density and metallicity of the corona gas simulataneously.
There, the best-fit gas density profile and the 68\% confidence-level
region of its parameters were determined in the frameworks of the commonly
used ``beta profile'' for the electron density,
\begin{equation}
n_{\rm e}(r) =n_{0} \left(1+\left(r/r_{\rm c} \right)^2   \right)^{-3
\beta/2},
\label{Eq:n_e}
\end{equation}
where $r$ is the distance to the Galactic center and $n_{0}$, $r_{c}$ and $
\beta$ are parameters.
The value of
$r_{\rm c} \lesssim 5$~kpc can hardly be constrained presently but does
not affect the density in the outer Galaxy
(we use $r_{\rm c}=3$~kpc in our numerical
calculations). At large distances $r$, the profile reduces to a simple
power-law falloff, $n_{e}\approx n_{0} r_{c}^{3
\beta} r^{-3\beta}$.
As it is stated in Ref.~\cite{ST}, however, these observational
constraints are related to $r \gtrsim 45$~kpc, while the gas profile at
smaller radial distances is poorly constrained. Given the lack of
observational constraints for the inner region of the corona (which,
however, is still outside the stellar halo!), we invoke the results of
computer simulations to describe the gas density profile there. To be
specific, we note that the simulated profile of Ref.~\cite{Gnedin} is well
approximated, at $r \lesssim 45$~kpc, by the same Eq.~(\ref{Eq:n_e}) with
\begin{equation}
\beta=0.843,
~~~~
\red
n_{0}=0.738~{\rm cm}^{-3}.
\black
\label{Eq:beta-Hooper}
\end{equation}
In our calculations, we use, for the gas density profile, the maximum of
two beta profiles, Eq.~(\ref{Eq:n_e}), one giving the fit to the simulated
profile of Ref.~\cite{Gnedin}, with parameters~(\ref{Eq:beta-Hooper}),
and another with
\begin{equation}
\beta=0.195,
~~~~
\red
n_{0}=0.00153~{\rm cm}^{-3},
\black
\label{Eq:beta-ST}
\end{equation}
determined as the best-fit profile in Ref.~\cite{ST}. Following
Ref.~\cite{astro-ph/0001142}, we assume that the gas density is cut at
$R_{\rm max}=250$~kpc. The resulting $n_{e}(r)$ is presented in
Fig.~\ref{fig:ngas}.
One can see that, indeed, the change between the two regimes takes place at
$r\sim 45$~kpc.

\section{Cosmic rays and their interactions}
\label{sec:CR}
Much less is known about the cosmic-ray densities and spectra in the
Galactic corona. As described in detail in Ref.~\cite{Gnedin}, this
region contains all cosmic-ray particles of relevant energies which escaped
from the Galactic disk throughout the Milky-Way lifetime. The propagation
of particles is determined by the diffusion coefficient $D$ which depends
in turn on the particle energy and location. The cosmic-ray escape from
the disk is \red non-spherical \black \cite{PtZ1, PtZ2},
but for purposes of the present work we assume
spherical symmetry;
the contribution of the very same physical processes
in the strongly asymmetric inner part, including the Galactic disk, have
been studied elsewhere \cite{Berezinsky, Ahlers, SemikozDisk}.

We will further consider the proton component of cosmic rays.
Within the approximation of spherical symmetry, the diffusion equation for
protons in the corona reads as
\begin{equation}
\frac{\partial j}{\partial t}(E,r) = D(E,r) \Delta j -c\sigma_{pp}(E)n_{cr}(r)j(E,r)+Q(E,r,t),
\label{Eq:diffusion}
\end{equation}
where $E$ is the cosmic-ray energy, $r$ is distance from Milky Way center,
$j=dn_{\rm CR}/dE$ is the cosmic-ray spectral density, $D$ is the
diffusion coefficient (we disregard its dependence on $r$ and assume the
Kolmogorov turbulence regime $D(E)=D_{0}(E/GeV)^{1/3}$), $t$ is time,
$\Delta$ denotes the radial part of the three-dimensional Laplace
operator, $\sigma_{pp}$ is p-p interaction cross section, $c$ is the speed
of light and $Q$ is the source term. We solve Eq.~(\ref{Eq:diffusion})
numerically to obtain the cosmic-ray concentration and spectrum for two
marginal values of the diffusion coefficient, $D_{0}=1.2 \times
10^{29}$~cm$^{2}$/s and $D_{0}=4 \times 10^{30}$~cm$^{2}$/s, used in
Ref.~\cite{Gnedin}. We assume the time-dependent source term, constant
within $r \le r_{\rm Q}=5$~kpc and zero at $r>r_{Q}$, having power-law
energy dependence with exponential cut-off. For the source evolution, we
use the same expression as in Ref.~\cite{Gnedin}, so
\begin{equation}
Q(E,r,t) \propto E^{-\alpha} \exp\left(-\frac{E}{E_{\rm max}}\right)
\Theta(r_{\rm Q}-r)\times
\begin{cases}
1 + t/(1\,{\rm Gyr}) & \text{if $t\leq{}2$ Gyr,} \\
3 & \text{if 2 Gyr $<t\leq{}$6 Gyr,} \\
3-0.5(t-6\,{\rm Gyr}) & \text{if 6 Gyr $<t\leq{}$10 Gyr,}
\end{cases}
\label{eq-source}
\end{equation}
where $\Theta$ is the step function.

The normalization of the source term is chosen in such a way that
the cosmic-ray flux at the solar location, $r\approx 8.5$~kpc, does not
exceed the observed one at all energies. We choose $\red \alpha
\black =2.4$ and $E_{\rm max}=10^{17}$~eV as our working model because this
saturates the observed spectrum (which is softer by $1/3$ because of the
Kolmogorov diffusion).

We comment briefly on other options in Sec.~\ref{sec:concl}.

\begin{figure}
\centerline{\includegraphics[width=0.95\columnwidth]{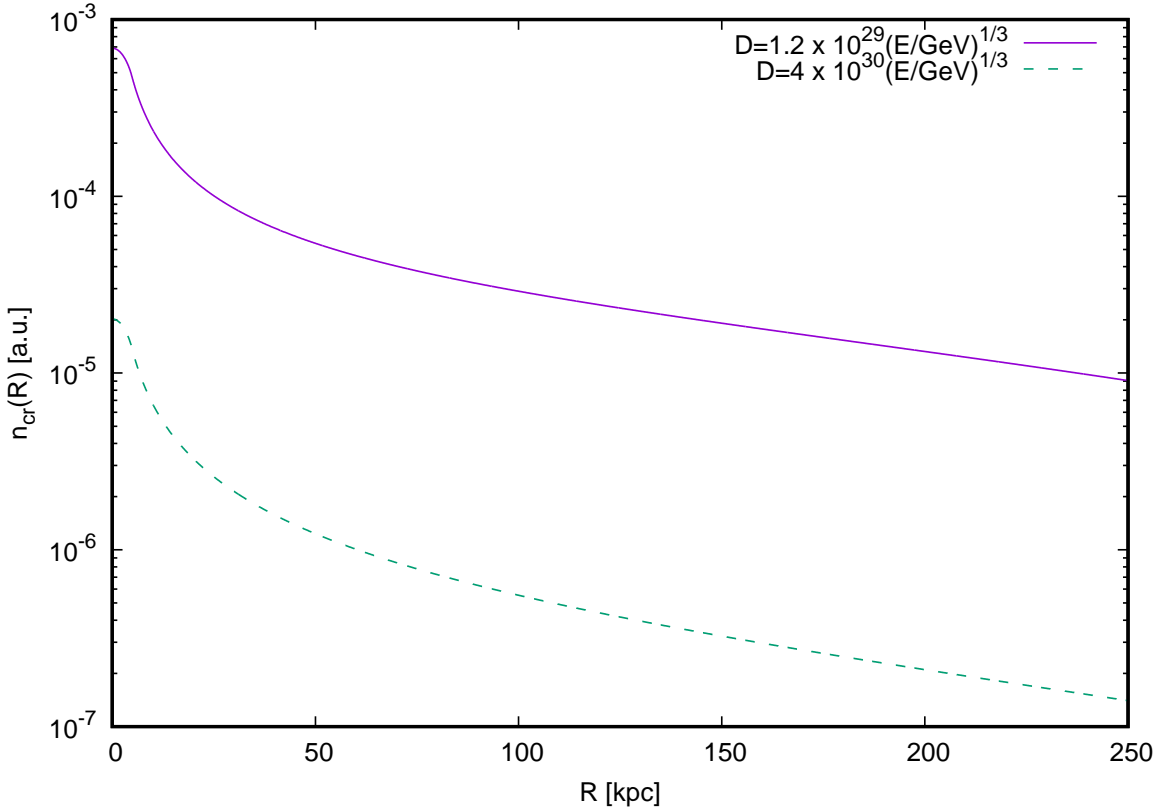}}
\caption{
\label{fig:D1}
The radial dependence of the cosmic-ray density, $n(r)$, in the Galactic
corona, obtained as a solution to the diffusion equation,
Eq.~(\ref{Eq:diffusion}), for two values of the diffusion coefficient
$D_{0}$ \oleginline{and for proton energy $E=0.1$~PeV}. }
\end{figure}

\begin{figure}
\centerline{\includegraphics[width=0.95\columnwidth]{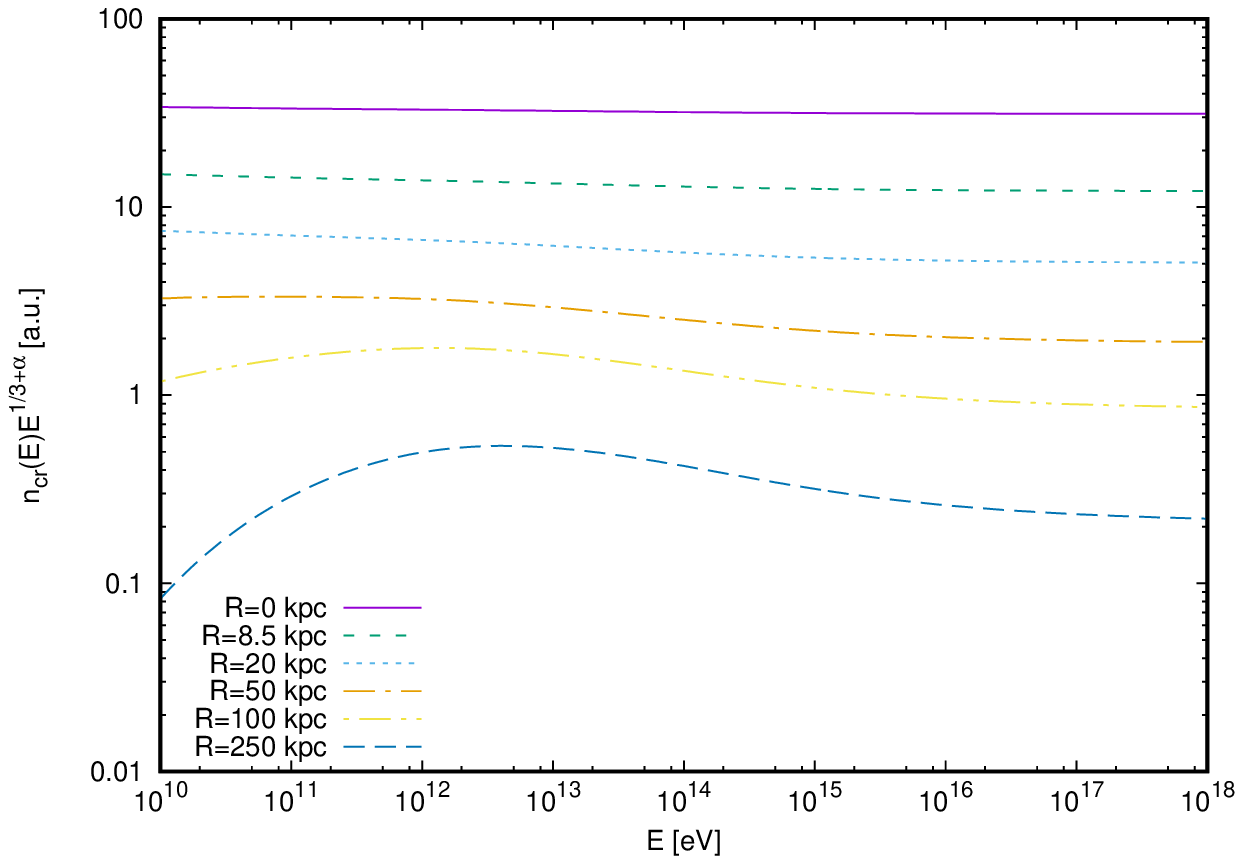}}
\caption{
\label{fig:D2}
The energy spectrum of the cosmic-ray density, $n(E)$, in the Galactic
corona, obtained as a solution to the diffusion equation,
Eq.~(\ref{Eq:diffusion}), at various galactocentric distances $r$. The
diffusion coefficient $D_{0}=1.2\times 10^{29}$~cm$^{2}$/s.  }
\end{figure}

Figures~\ref{fig:D1}, \ref{fig:D2}
present the resulting cosmic-ray density in the corona obtained from the
solution of Eq.~(\ref{Eq:diffusion}). These numerical results are in a
good agreement with general expectations.
The
cosmic-ray density in the corona should fall off like $1/r^{A}$, where
$1 \le A \le 2$ (the value $A=2$ corresponds to the free escape with
$D=\infty$ while $A=1$ corresponds to a constant $D$). The spectrum of
cosmic rays is harder, compared to the injected one, with the difference of
power-law spectral indices of $1/3$, assuming the Kolmogorov turbulence.

To calculate the fluxes of photons and neutrino we use the open-source
numerical code~\cite{Kalashev:2014xna} for solving transport equations in
one dimension. The code~\cite{Kalashev:2014xna} has been extended to
include $pp$
interactions as follows:
The inelastic cross sections $\sigma_{\rm inel}$ of CR nuclei on gas were
calculated with QGSJET-II-04~\cite{qgs2-4}.  For the spectrum of secondary
photons and neutrinos produced in $pp$ interactions, differential cross
sections tabulated from QGSJET-II-04 were used~\cite{fragpp}.

The Milky-Way corona contribution is direction-dependent because of
the non-central position of the Sun in the Galaxy. We calculate the flux
from each direction by solving the transport equation for propagation along
the corresponding straight line with the source term proportional to
$n_g(r)n_p(r)$,
\begin{equation}
Q_i(E_i,r) = n_g(r)\int dE_p n_p(E_p,r)\frac{d\sigma_{pp}}{dE_i}(E_p),  \,
i=\nu,\gamma.
\end{equation}
We include the $e^{+}e^{-}$ pair production term to the transport equation
to take into account that $\gamma$-rays with $E\gsim 100$~TeV are
subjected to attenuation on CMB photons. Note that their interaction with
infra-red background is negligible on the scales of hundreds of kpc.

\section{Contribution of other galaxies}
\label{sec:EG}
It has been pointed out in Ref.~\cite{Gnedin} that the contribution of
coronae of other galaxies to the diffuse gamma-ray background is
comparable to that of our own Galaxy. Here we assume that properties of
galactic coronae throughout the Universe are similar and estimate
their contribution to the observed gamma-ray and neutrino fluxes. The
total amount of cosmic rays in coronae is determined by the
evolution of galaxies. Indeed, circumgalactic gas structures similar to
the Milky Way have been recently found in other galaxies, see e.g.\
Refs.~\cite{Postnov, other-halo}. Assuming that all cosmic rays, which had
been produced in a galaxy and subsequently escaped from it, are now
contained within its virial radius (this is true~\cite{Gnedin} for the
Milky Way) and that the cosmic rays are accelerated in some stellar
processes, related e.g.\ to supernova explosions, one concludes that the
amount of cosmic rays is proportional to the total stellar mass in a
galaxy. Note that the amount of cosmic rays in a disk is often assumed to
be proportional to the star formation rate which is the time derivative of
the total mass. This is because cosmic-ray particles do not stay in a disk
for long.

Therefore, to calculate the photon, $\gamma$, and neutrino, $\nu$, fluxes
from coronae of remote galaxies, we solve the transport equations
with the source term proportional to the total stellar mass density
$\rho^{\star}(z)$ at the redshift $z$,
\begin{equation}
Q_i(E_i,z) \propto \rho^{\star}(z)\frac{d\sigma_{pp}}{dE_i}(E_p), \,
i=\nu,\gamma,
\end{equation}
and include extra term for $\gamma$ interactions with extragalactic
background light (EBL) into the transport equation. For the EBL, we employ
the baseline model of Ref.~\cite{EBL}.

We use the redshift-dependent total stellar mass function of galaxies
$\rho^{\star}(z)$  given in Ref.~\cite{SMF} to parametrise the relative
normalization of injected cosmic-ray fluxes as a function of redshift. All
of the contributions are summed up (for photons, with the account of
absorption) and, knowing the total stellar mass of the Milky Way (see
e.g.\ Ref.~\cite{SM-MW}), we relate the overall normalization of the
extragalactic contribution to the flux from the Milky-Way corona as
follows.

To obtain the relative extragalactic contribution,
we use the fact that the $pp$ cross section $\sigma$ and the average
number $\kappa$ of neutrinos produced in a single $pp$ interaction
moderately depend on the cosmic-ray proton energy $E_p$ for sufficiently
high $E_p \gsim$TeV.
For this reason, the secondary neutrino spectrum roughly follows the
cosmic-ray proton spectrum for which we assumed the power-law dependence.
Therefore, we omit the energy dependence in the following equations.

Since the amount
of cosmic rays in a galactic corona scales with the total stellar mass
of the galaxy $M^{\star}$, one may write, \red for the
cosmic-ray density,   \black
$$
\red
n_{\rm CR}(r)=\frac{M^{\star}}{M^{\star}_{\rm MW}}\bar n_{\rm
CR}(r),
\black
$$
where the function $\bar n_{\rm CR}(r)$ is universal. The
Milky-Way contribution to the neutrino flux from the
Galactic anticenter direction is then
$$
j_{\rm MW}=
\red
\frac{c}{4\pi} \kappa\sigma
\black
\, \int_{R_{\odot}}^{R_{\rm max}}  \bar n_{\rm
CR} n_g \, dr
\equiv
\frac{c}{4\pi} \kappa\sigma  M^{\star}_{\rm MW}
I_{\rm MW},
$$
where $R_{\odot}=8.5$~kpc is the distance from the Sun to the
Galactic center and  $R_{\rm max}=250$~kpc is the corona radius.

The extragalactic contribution is given by the integral over the volume of
the Universe,
$$
j_{\rm EG}=\frac{c}{4\pi} \kappa\sigma
\int dV\,\frac{1}{4\pi d^2(1+z)} \red \frac{\rho^\star}{M^{\star}_{\rm
MW}} \black \, \int_0^{R_{\rm max}} \bar n_{\rm CR} n_g 4\pi r^2 \, dr,
$$
where $\rho^{\star}$ is the total stellar mass comoving density, $d$ is the
comoving distance and the $(1+z)$ denominator takes into
account redshifting of the time interval. We denote
$$
I_z \equiv \int dV \, \frac{1}{4\pi d^2} \rho^\star
$$
and
$$
I_V=\int_0^{R_{\rm max}}   \bar n_{\rm CR} n_g r^2 \, dr
$$
to obtain
$$
\xi \equiv \frac{j_{\rm MW}}{j_{\rm EG}}
=
\frac{M^\star_{\rm MW} I_{\rm MW}}{4 \pi c I_z I_V}.
$$
One should take into account the redshift difference between the observed
and emitted energies, $E'=(1+z)E$, to obtain, for a power-law spectrum,
$dN/dE \propto E^{-\alpha}$,
$$
I_z = {1 \over H_0}
\int\limits_0^{z_{\rm max}}
\frac{\rho^\star (z) (1+z)^{-\alpha}}{\sqrt{(1+z)^3 \Omega_M +
\Omega_\Lambda}}\, dz,
$$
where $H_{0}$ is the Hubble constant, $\Omega_M$ and $\Omega_\Lambda$ are
cosmological matter and vacuum energy densities, respectively.

The total stellar mass density $\rho^\star (z)$ may be determined from the
total stellar mass function,
$$
\Phi(M^\star,z)=\frac{dN}{dV d \lg M^\star},
$$
as
$$
\rho^\star =\frac{1}{\ln 10} \int\Phi\, dM^\star,
$$
and the function $\Phi$ we use is presented in Ref.~\cite{SMF} (Model~3).
The value of the total stellar mass in the Milky Way, $M^{\star}_{\rm MW} =
\left(6.43 \pm 0.63 \right) \times 10^{10} M_{\odot}$, is taken from
Ref.~\cite{SM-MW}. The value of $\xi$ depends on the assumed spectral
index $\alpha$ but in all realistic cases $\xi < 1$. This calculation
neglects the absorption and the cut-off of the proton spectrum
and therefore is valid for neutrino fluxes far away from the cut-off
region; the propagation effects for photons are taken into account
numerically as it is described above.

\section{Results and discussion}
\label{sec:concl}
\begin{figure}
\centerline{\includegraphics[width=0.95\columnwidth]{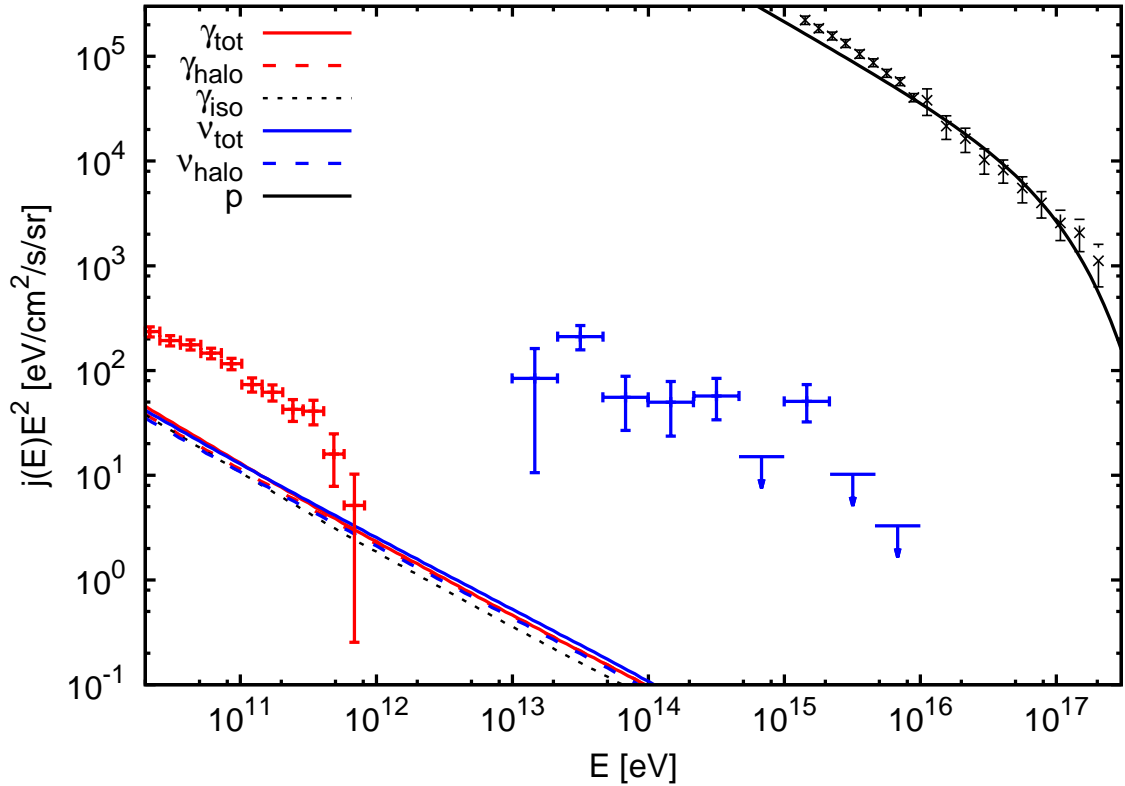}}
\caption{
\label{fig:spec}
Diffuse neutrino (blue lines) and gamma-ray (red
lines) spectra predicted by the model described in the text
($\alpha=2.4$, $E_{\rm max}=10^{17}$~eV, $D_{0}=1.2\times
10^{29}$~cm$^{2}$/s). The Milky-Way corona contribution is shown by
\oleginline{dashed} lines, the total flux from coronae of all galaxies,
including the Milky Way, is shown by solid lines. The isotropic part of
gamma-ray flux is shown by the black dotted line. The blue error bars:
IceCube astrophysical neutrino flux \cite{IceCubePoints}. The red
error bars: FERMI-LAT extragalactic gamma background
flux~\cite{FERMIbckgr} (the upper flux level is shown allowed by galactic
foreground model uncertainty). Crosses: total cosmic-ray flux measured by
KASCADE and KASCADE-Grande~\cite{CRpoints}.}
\end{figure}

\begin{figure}
\centerline{\includegraphics[width=0.95\columnwidth]{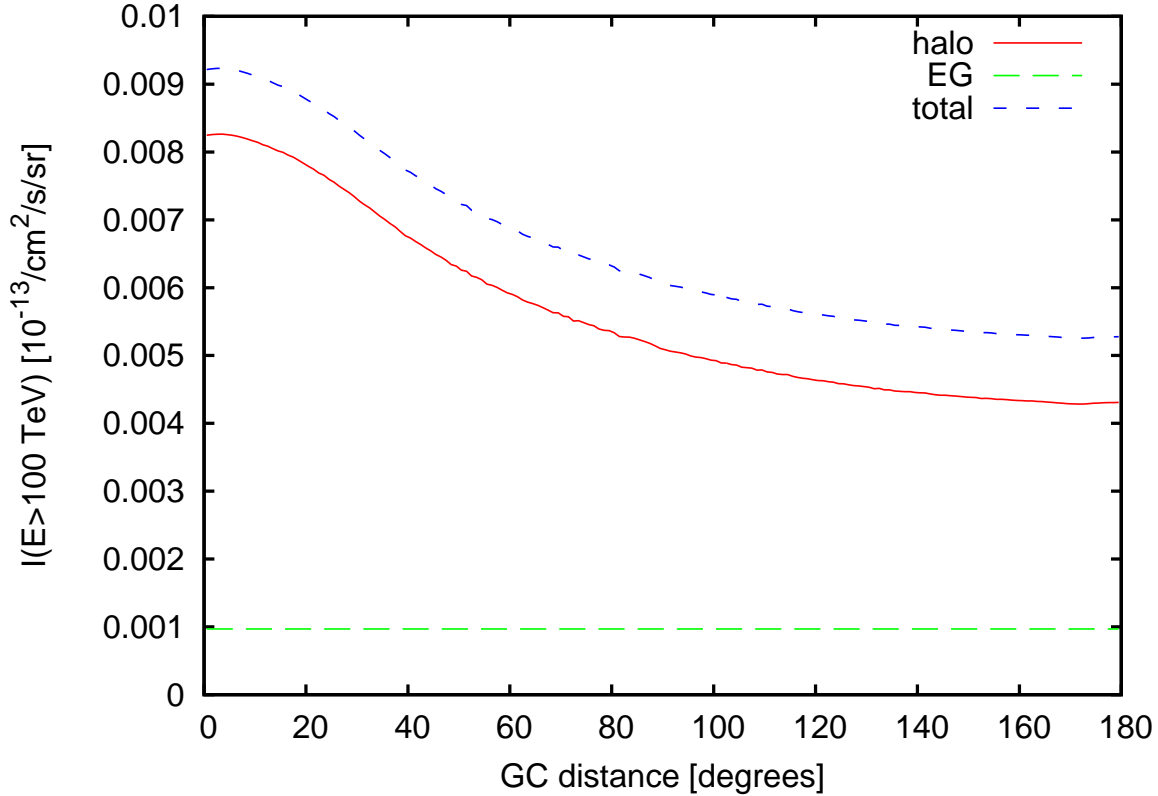}}
\caption{
\label{fig:aniso}
Galactic dipole anisotropy of the diffuse fluxes.
}
\end{figure}

Results of the calculation are presented in Fig.~\ref{fig:spec},
where,
for our baseline scenario (cosmic-ray injection spectrum with $\alpha
=2.4$ and   $E_{\rm max}=10^{17}$~eV) we show the spectra of secondary
gamma rays and neutrinos from cosmic-ray interactions with the
circumgalactic gas. One can see that this contribution to the IceCube
astrophysical neutrino flux is almost negligible and does not exceed 1\%.
At the same time, these interactions provide a sizeable contribution to
the diffuse gamma-ray background at FERMI-LAT energies, in accordance with
Ref.~\cite{Gnedin}. Due to the non-central position of the Sun in the
Galaxy, this contribution is not isotropic but follows a dipole-like
pattern with respect to the Galactic center, see Fig.~\ref{fig:aniso}.

\red
One may compare the expected neutrino flux from cosmic-ray interactions in
the corona with a similar contribution from the Galactic disk. For the
conventional assumptions about the cosmic-ray spectrum in the Galaxy,
which correspond to our $\alpha=2.4$ injection spectrum, the disk
contribution was estimated in Ref.~\cite{Ahlers} to be about $(4-8)\%$ of
the IceCube neutrino flux, significantly exceeding the corona
contribution. Note that the Milky-Way disk contribution to the neutrino
flux should exhibit a clear Galactic-plane anisotropy in the arrival
directions, very different from the dipole anisotropy from the corona.
While there exist some claims of the Galactic-plane
excess~\cite{SemikozDisk}, inclusion of the muon track events in the
analysis makes them insignificant~\cite{ST-dipole}. \black

One may attempt to discuss possible variations in the model which might
enlarge the neutrino flux. The largest uncertainty in our calculations is
related to the cosmic-ray concentration and spectrum in the corona. Let us
discuss the effect of its possible variations.

\begin{figure}
\centerline{\includegraphics[width=0.95\columnwidth]{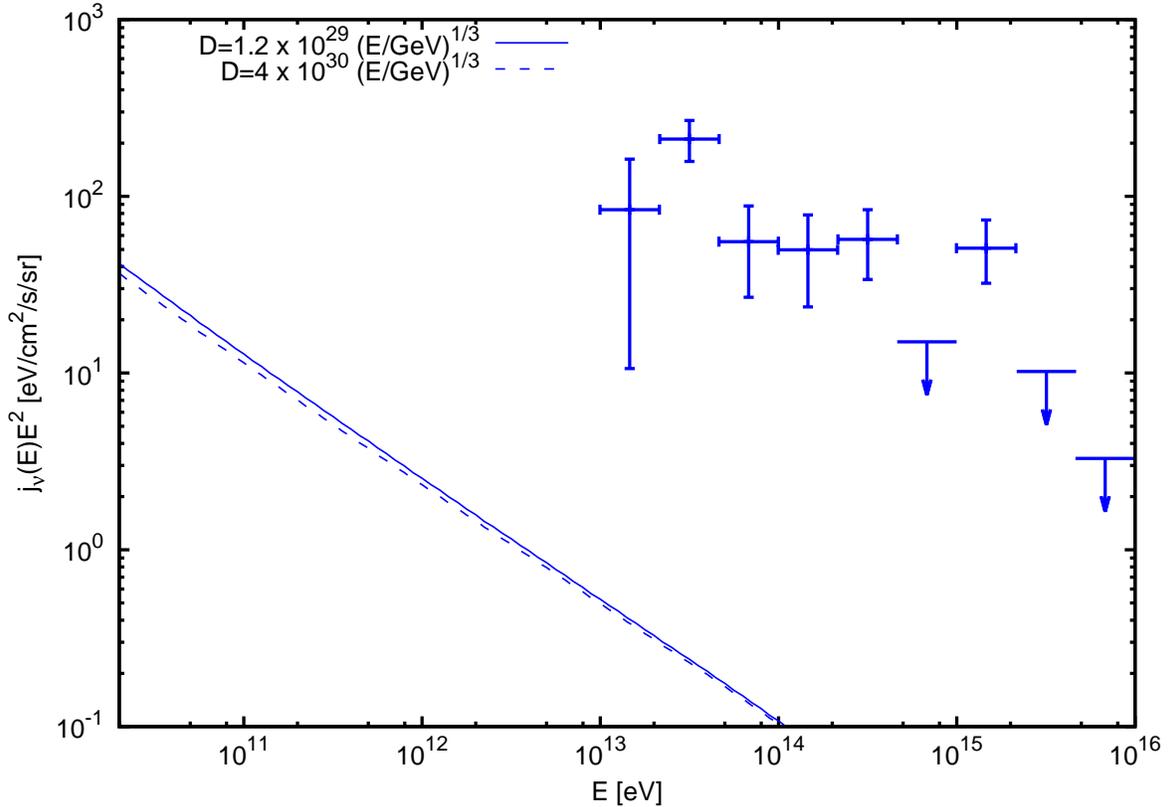}}
\caption{
\label{fig:D-compare}
The diffuse neutrino flux calculated for two values of the diffusion
coefficient. Data points represent the IceCube astrophysical neutrino flux.
}
\end{figure}

\textit{(i) Variations in the diffusion coefficient.} The coefficient $D$
depends strongly on the value of the magnetic field in the outer halo,
which is poorly constrained.  The solution to the diffusion equation with
fixed source normalization depends on $D$, as it is obvious from
Fig.~\ref{fig:D1}. However, since we choose the normalization of the source
term $Q(E,r,t)$ to touch the observed cosmic-ray spectrum, the variations
in $D$ by an order of magnitude do not change the result qualitatively,
see Fig.~\ref{fig:D-compare}.

\red In our calculations, we have normalized the cosmic-ray density in the
corona in such a way that the observed cosmic-ray spectrum at the Earth
location is never exceeded. The diffusion in the Galactic disk is likely
slower than in the corona, and additional amount of cosmic rays confined
in the disk contribute to the observed spectrum. Account of this effect
would further reduce the normalization of the cosmic-ray density in the
corona, and hence the diffuse gamma-ray and neutrino fluxes we discuss
here. \black

\begin{figure}
\centerline{\includegraphics[width=0.95\columnwidth]{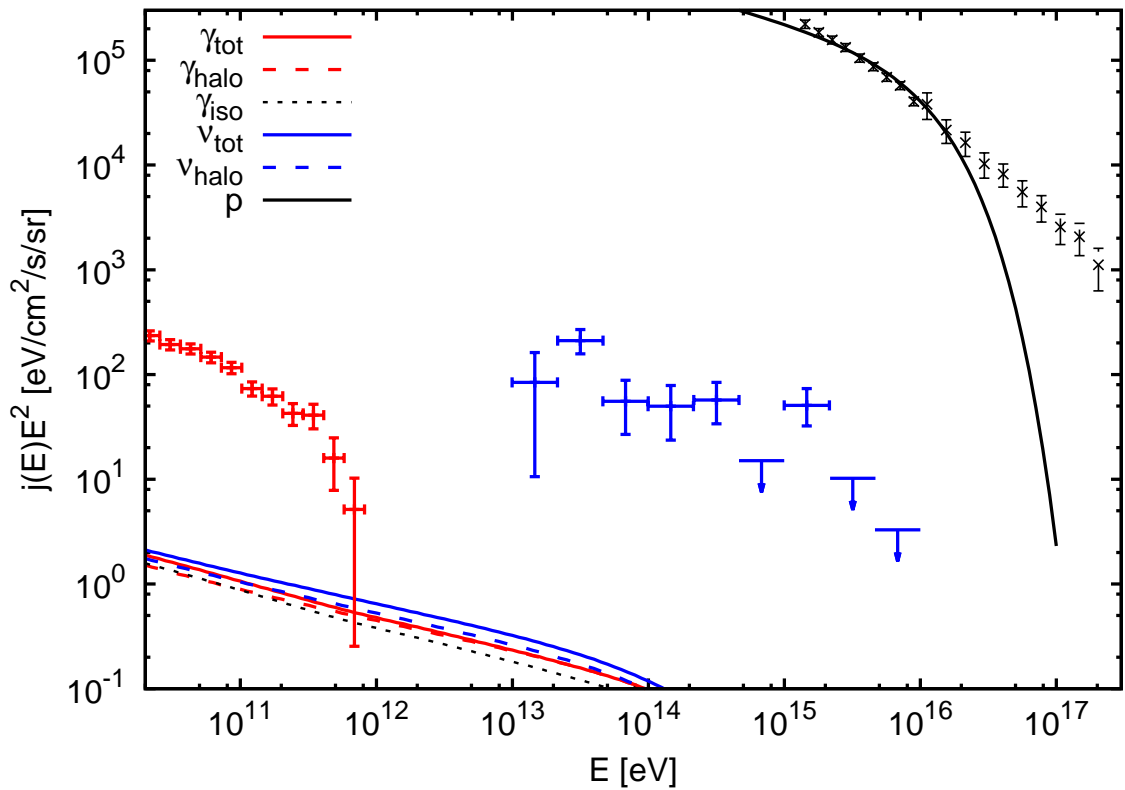}}
\caption{
\label{fig:specHard}
The same as in Fig.~\ref{fig:spec} but for the hard-spectrum model
($\alpha=2.0$, $E_{\rm max}=10^{16}$~eV, $D_{0}=1.2\times
10^{29}$~cm$^{2}$/s).
}
\end{figure}

\textit{(ii) Variations in the cosmic-ray spectral shape.}
The spectrum of IceCube astrophysical neutrinos looks harder than the one
we obtained in our simulations with $\alpha =2.4$. Examination of
Fig.~\ref{fig:spec} suggests that taking a harder spectrum of the original
cosmic rays might result in a considerably larger
neutrino flux. Note that $\alpha =2$ is predicted by the usual
second-order Fermi acceleration, and some recent studies, see e.g.\
Ref.~\cite{Hard}, suggest that the soft, $\alpha \approx 2.7$, cosmic-ray
spectrum observed at the Earth is attributed to the effect of a nearby
soft source. However, in the frameworks of our approach, hardening of the
injection spectrum results in the suppression of its normalization,
required in order not to overshoot the observable cosmic-ray spectrum at
high energies. Figure~\ref{fig:specHard}
presents our results for the injected cosmic-ray spectrum with
$\alpha=2.0$ and $E_{\rm max}=10^{16}$~eV. Even in this extreme case, the
neutrino flux cannot contribute to the IceCube flux significantly.
\red Note that the disk contribution enhances under these
assumptions~\cite{SemikozDisk}. \black

\textit{(iii) Variations in the cosmic-ray injection history.}
A very important assumption for our calculations is the choice of the time
dependence of the source
term (\ref{eq-source}) in the cosmic-ray diffusion equation.
If cosmic rays are accelerated, e.g., in supernova shocks, then the source
term should be proportional to the time-dependent star formation rate in
the Galaxy, which motivates the time dependence used in
Eq.~(\ref{eq-source}). This leads to a temporal, factor of $\sim 3$,
enhancement in the injected spectrum, which does not affect
our conclusions qualitatively.

However, other sources of cosmic-ray protons might work at various stages
of the Milky-Way history. Temporary activity of the Galactic Center could,
in principle, result in an additional contribution of energetic protons
which, by now, may find themselves already in the corona. Both the shape
and the normalization of their spectrum are hardly constrained by the
present-day spectrum observed at the Earth. The interactions of
these particles with the circumgalactic gas may lead to a considerable
enhancement of secondary neutrino and gamma-ray fluxes at high energies.
In this case, diffuse gamma rays would provide an important constraint on
the model~\cite{Kohta, OK-ST-gamma}. We will address this interesting
possibility in a forthcoming work~\cite{OK-ST-progress}.

\section*{Acknowledgements}
We are indebted to A.~Bykov, M.~Pshirkov, V.~Ptuskin and
G.~Rubtsov for useful and stimulating discussions. OK thanks
Aspen Center for Physics for hospitality and ST thanks CERN
(TH department) for hospitality at the initial stages of this work. This
work is supported by the Russian Science Foundation, grant 14-12-01340.
Numerical calculations have been performed at the computer cluster of the
Theoretical Physics Department, Institute for Nuclear Research of the
Russian Academy of Sciences.

\end{document}